\documentclass[aps,prl,twocolumn,letterpaper,showpacs]{revtex4}

\usepackage{times}
\usepackage[T1]{fontenc}
\usepackage[latin1]{inputenc}
\usepackage[psamsfonts]{amssymb}
\usepackage{graphicx}
\usepackage{amsbsy}
\usepackage{amsmath} 

\begin{document}

\title{Phase coupling estimation from multivariate phase statistics}

\author{Charles F.~Cadieu}
\author{Kilian Koepsell}
\affiliation{
Redwood Center for Theoretical Neuroscience \\
Helen Wills Neuroscience Institute \\
University of California, Berkeley \\
Berkeley, CA 94720\\
\texttt{\{cadieu,kilian\}@berkeley.edu}
}


\begin{abstract}
  Coupled oscillators are prevalent throughout the physical world. Dynamical
  system formulations of weakly coupled oscillator systems have proven
  effective at capturing the properties of real-world systems. However, these
  formulations usually deal with the `forward problem': simulating a system
  from known coupling parameters. Here we provide a solution to the `inverse
  problem': determining the coupling parameters from measurements.  Starting
  from the dynamic equations of a system of coupled phase oscillators, given
  by a nonlinear Langevin equation, we derive the corresponding equilibrium
  distribution. This formulation leads us to the maximum entropy distribution
  that captures pair-wise phase relationships. To solve the inverse problem
  for this distribution, we derive a closed form solution for estimating the
  phase coupling parameters from observed phase statistics. Through
  simulations, we show that the algorithm performs well in high dimensions
  (d$=$100) and in cases with limited data (as few as 100 samples per
  dimension). Because the distribution serves as the unique maximum entropy
  solution for pairwise phase statistics, the distribution and estimation
  technique can be broadly applied to phase coupling estimation in any system
  of phase oscillators.
\end{abstract}









\maketitle

Many complex natural phenomena can be modeled as networks of coupled
oscillators. Examples can be drawn from the physical, chemical, and biological
world. Oscillator models have been effective at describing the dynamics of
coupled pendula, coupled Josephson junctions, reaction diffusion systems,
circadian rhythms, oscillating neural networks, and even the coupling of
firefly luminescence (see e.g.~\cite{Buzsaki2006, Kuramoto1984, Mirollo1990,
  Strogatz2003, Winfree2001}).

In many systems, coupling topology and the strength of interaction between
network elements is of central scientific interest. However, network coupling
often can not be measured directly and must be inferred from
measurements. Therefore the inverse problem, or inferring network coupling
from measurements, is of central importance.

In statistical mechanics, the inverse problem is typically solved by proposing
a probability distribution and estimating the distribution's parameters from
measurements. A natural choice for the estimation, a highly under-determined
problem, is given by the unique maximum entropy distribution that reproduces
the statistics of the measurements~\cite{Jaynes1957}. A number of such
distributions and estimation techniques are used throughout the science and
engineering communities. In the real-valued case the multivariate Gaussian
distribution, and in the binary case the Ising model, serve as widely used
multivariate maximum entropy distributions consistent with second order
statistics. Each of these cases has well known estimation techniques for
inferring the distribution's parameters from observations. The availability of
these techniques has led to a number of applications, e.g.\ the Ising model
and its corresponding estimation techniques have been used to infer the
coupling in networks of retinal ganglion
cells~\cite{Shlens2006,Schneidman2006}. However, for the phase variables that
are of interest in networks of oscillators there has been little work on
providing a corresponding multivariate probabilistic distribution, or deriving
estimation techniques to infer the distribution's parameters from data.

In this Letter, we provide a solution to the inverse problem for systems of
coupled phase oscillators. We begin by presenting the Langevin dynamics for a
generalized form of the Kuramoto model of coupled phase oscillators. Solving
for the equilibrium distribution yields a multivariate probability
distribution of coupled phase variables. This probabilistic formalism allows
us to derive a novel estimation technique for the coupling terms from phase
variable measurements. We show that this technique performs robustly with
limited data and in high dimensions.

Consider a network of $d$ identical coupled oscillators with intrinsic
frequency $\omega$. In the limit of weak coupling, the amplitude of the
oscillators can be assumed to be constant and the equations of motion can be
formulated in terms of $d$ phase variables $0\!\le\!\theta_i\!<\!2\pi$,
$i\!=\!1,\ldots,d$. A popular choice for the dynamics of such a system is
given by the Kuramoto model~\cite{Kuramoto1984}, which has constant coupling
between oscillators. We can generalize this model to include inhomogeneous
couplings and inhomogeneous phase offsets between oscillators. The dynamic
equation is then given by
\begin{equation}
  \frac{\partial}{\partial t}\theta_i(t)\!=\omega -
  \kappa_{ij}\sum_{j=1}^d\!\sin(\theta_i(t)\!-\!\theta_j(t)\!-\!\mu_{ij}) + 
  \Gamma_i(t),
\label{eq:kuramoto}
\end{equation}
where $\kappa_{ij}$ is the coupling strength and $\mu_{ij}$ is the preferred
phase between two oscillators $i$ and $j$. We only consider the case of
symmetric coupling ($\kappa_{ij}\!=\!\kappa_{ji}$,
$\mu_{ij}\!=\!-\mu_{ji}$). The noise fluctuations, $\Gamma_i(t)$, are zero
mean Gaussian distributed with $\delta$ covariance functions and variance
$\beta^{-1}$, corresponding to the temperature of the system:
\begin{equation}
  \langle \Gamma_i(t) \rangle = 0, 
  \quad \langle \Gamma_i(t) \Gamma_j(t') \rangle = 
  2\,\beta^{-1}\delta_{ij}\delta(t-t')\,.
\label{eq:noise}
\end{equation}

The equations of motion~(\ref{eq:kuramoto}) for our system of coupled
oscillators can be considered as a nonlinear Langevin equation describing
Brownian motion on a $d$-torus in the presence of the potential
$E(\boldsymbol{\theta})$ given by
\begin{equation}
  E(\boldsymbol{\theta}) = -\frac{1}{2} \sum_{i,j=1}^d 
  \kappa_{ij}\cos(\theta_i-\theta_j-\mu_{ij}) \,,
\label{eq:energy}
\end{equation}
where $\boldsymbol{\theta}$ is now a $d$-dimensional vector with components
$\theta_i$. Note that by applying the transformation
$\tilde{\theta}_i(t)\!=\!\theta_i(t)-\omega t$ to~(\ref{eq:kuramoto}) we can
assume $\omega\!=\!0$ without loss of generality.

By changing the coordinates from the angular representation,
$\boldsymbol{\theta}$, to the complex representation,
$\{\mathbf{x}\!\in\!\mathbb{C}^d\,\boldsymbol{|}\,|x_k|\!=\!1\}$ with
components $x_k=e^{i\theta_k}$, we can rewrite eq.~(\ref{eq:energy}) more
compactly as the (real-valued) quadratic Hermitian form:
\begin{equation}
  E(\mathbf{x}) = 
  -{\textstyle \frac{1}{2}}\, \mathbf{x}^\dagger\mathbf{K}\mathbf{x}\,,
\label{eq:energy-hermitean}
\end{equation}
where $\mathbf{K}$ is a $d\!\times\!d$ Hermitian matrix with elements $K_{jk}
= \kappa_{jk}e^{i \mu_{jk}}$. This energy function~(\ref{eq:energy}) is
closely related to the XY-model, which only has homogeneous nearest neighbor
couplings ($k_{ij}$ = const.) and no phase offsets ($\mu_{ij}$=0). This
generalization is analogous to the extension of the homogeneous Ising model to
spin glasses.

It is known (see e.g.~\cite{Risken1989a}) that the probability density
$p(\boldsymbol{\theta}(t),t)$ of a system governed by Langevin dynamics
evolves according to the Fokker-Planck equation
\begin{equation}
  \frac{\partial p(\boldsymbol{\theta},t)}{\partial t}  = 
  -\sum_i\frac{\partial}{\partial \theta_i} D_i p(\boldsymbol{\theta},t)
  +\sum_{ij}\frac{\partial^2}{\partial\theta_i \partial\theta_j} 
  D_{ij} p(\boldsymbol{\theta},t) \,,
\label{eq:fokker-planck}
\end{equation}
with drift and diffusion coefficients given by
\begin{equation}
  D_i = -\frac{\partial E(\boldsymbol{\theta})}{\partial\theta_i}, \quad
  D_{ij} = \beta^{-1} \delta_{ij}\,.
\label{eq:current}
\end{equation}

Since the drift coefficient $D_i$ is a gradient field and the diffusion
coefficient $D_{ij}$ is constant, we can solve the Fokker-Planck
equation~(\ref{eq:fokker-planck}) for the stationary solution in closed form
and obtain
\begin{equation}
  p(\boldsymbol{\theta}) =
   \frac{1}{Z(\mathbf{K})} \exp[-\beta E(\boldsymbol{\theta})]
\label{eq:distribution}
\end{equation}
with the the energy function $E(\boldsymbol{\theta})$ given
by~(\ref{eq:energy}), and partition function $Z(\mathbf{K})$. 

We wish to solve the inverse problem for the general case of coupled phase
oscillators in equation~(\ref{eq:distribution}). Stated explicitly, the
problem is to infer the distribution's parameters (coupling terms
$\kappa_{ij}$ and phase offsets $\mu_{ij}$) from measurements of the network's
state, $\boldsymbol{\theta}$. 

The inverse problem is typically solved by following a maximum likelihood
estimation procedure.  Given the likelihood function $q(\boldsymbol{\theta})$
and the data distribution $p(\boldsymbol{\theta})$, the maximum likelihood of
the observed data with respect to the distribution parameters can be computed
by setting the derivative of the log-likelihood function to zero,
\begin{equation}
\label{eq:likelihood}
\frac{\partial
  \langle \log q(\boldsymbol{\theta})
  \rangle_{p(\boldsymbol{\theta})}}{\partial K_{ij}}
  \!\!= -\bigg\langle \frac{\partial E}{\partial K_{ij}} 
     \bigg\rangle_{p(\boldsymbol{\theta})}
    \!\!+\bigg\langle \frac{\partial E}{\partial K_{ij}}
     \bigg\rangle_{q(\boldsymbol{\theta})}
  \!\!\stackrel{!}{=}\, 0\,,
\end{equation}
where $\langle\ldots\rangle_{q(\boldsymbol{\theta})}$ denotes the expectation
value taken over the distribution $q(\boldsymbol{\theta})$. In our situation,
a closed form solution to this equation does not exist. However, we can find a
solution by iteratively descending the gradient. This procedure has a number
of drawbacks: the procedure is inherently iterative, estimating the
expectation under the estimated distribution $q(\boldsymbol{\theta})$ in
equation~(\ref{eq:likelihood}) involves a computationally expensive sampling
procedure, and the sampling procedure may suffer from a variety of problems
due to the landscape of the energy function.

To avoid the pitfalls of the maximum likelihood approach, we now derive a
closed form solution to the inverse problem for phase coupled oscillators. We
use the score-matching method introduced by
Hyvarinen~\cite{Hyvarinen2005,Hyvarinen2007}. Score-matching allows the
fitting of probability distributions of the exponential form for real-valued
data without computation of the normalization constant $Z$. If the energy
depends linearly on the distribution parameters, the solution can be computed
in closed form by setting the derivative of the score function to
zero~\cite{Hyvarinen2007}.

We follow this approach to estimate the distribution parameters for our
distribution in equation~(\ref{eq:distribution}). The score matching estimator
of $\mathbf{K}$ is given by $ \widehat{\mathbf{K}} = \arg \min_\mathbf{K}
J_{\mathrm{SM}}(\mathbf{K}) $ and the score function
$J_{\mathrm{SM}}(\mathbf{K})$ is given by
\[
J_{\mathrm{SM}}(\mathbf{K}) = \left\langle \textstyle{\frac{1}{2}}
[\nabla_{\boldsymbol{\theta}}E(\boldsymbol{\theta})]
[\nabla_{\boldsymbol{\theta}}E(\boldsymbol{\theta})]^T
- \nabla_{\boldsymbol{\theta}}^2E(\boldsymbol{\theta})
\right\rangle
\]
with the expectation value, $\langle\ldots\rangle$, taken over the data
distribution. Using the quadratic form of the energy in~(\ref{eq:distribution})
and the Jacobian $D_{ij}:=\partial x_i/\partial\theta_j$, we compute
\begin{eqnarray}
\nabla_{\boldsymbol{\theta}}E
&=&-\textstyle{\frac{1}{2}} \mathbf{x}^\dagger\mathbf{K}\mathbf{D}
   -\textstyle{\frac{1}{2}} \mathbf{D}^\dagger\mathbf{K}\mathbf{x}\nonumber\\
   \nabla_{\boldsymbol{\theta}}^2 E
&=& \mathbf{x}^\dagger\mathbf{K}\mathbf{x} 
   -\mathrm{Tr} (\mathbf{D}^\dagger\mathbf{K}\mathbf{D}) =
   -2\,E\nonumber\\
{}[\nabla_{\boldsymbol{\theta}}E]
{}[\nabla_{\boldsymbol{\theta}}E]^T
&=& \textstyle{\frac{1}{2}} \mathbf{x}^\dagger\mathbf{KK}\mathbf{x}
   +\textstyle{\frac{1}{4}} \mathbf{x}^\dagger\mathbf{KD}\mathbf{D}^T\mathbf{K}^T
\mathbf{x}^* \nonumber\\
&&   +\textstyle{\frac{1}{4}} \mathbf{x}^T\mathbf{K}^*\mathbf{D}^*\mathbf{D}^\dagger
\mathbf{K}\mathbf{x}
\nonumber
\end{eqnarray}
The estimator $\widehat{\mathbf{K}}$ is computed by setting the derivative of
the score function $\partial/\partial K_{ij}\,J_{\mathrm{SM}}(\mathbf{K})$ to
zero. This produces a system of linear equations,
\begin{equation}
\sum_{k,l=1}^{d} (
  \delta_{jl}C_{ik}
 +\delta_{ik}C_{lj}
 -\delta_{jk}C_{iljk}
 -\delta_{il}C_{iljk}
) \,\widehat K_{kl} 
= 4\,C_{ij},
\label{eq:linearsystem}
\end{equation}
where the expectation values are defined as $C_{ij}\!=\!\langle x_ix_k^*
\rangle$ and $C_{ijkl}\!=\!\langle x_ix_jx_k^*x_l^*\rangle$. We can solve this
system of linear equations using standard techniques.

\begin{figure}[ht]
\begin{center}
\includegraphics[width=8.5cm]{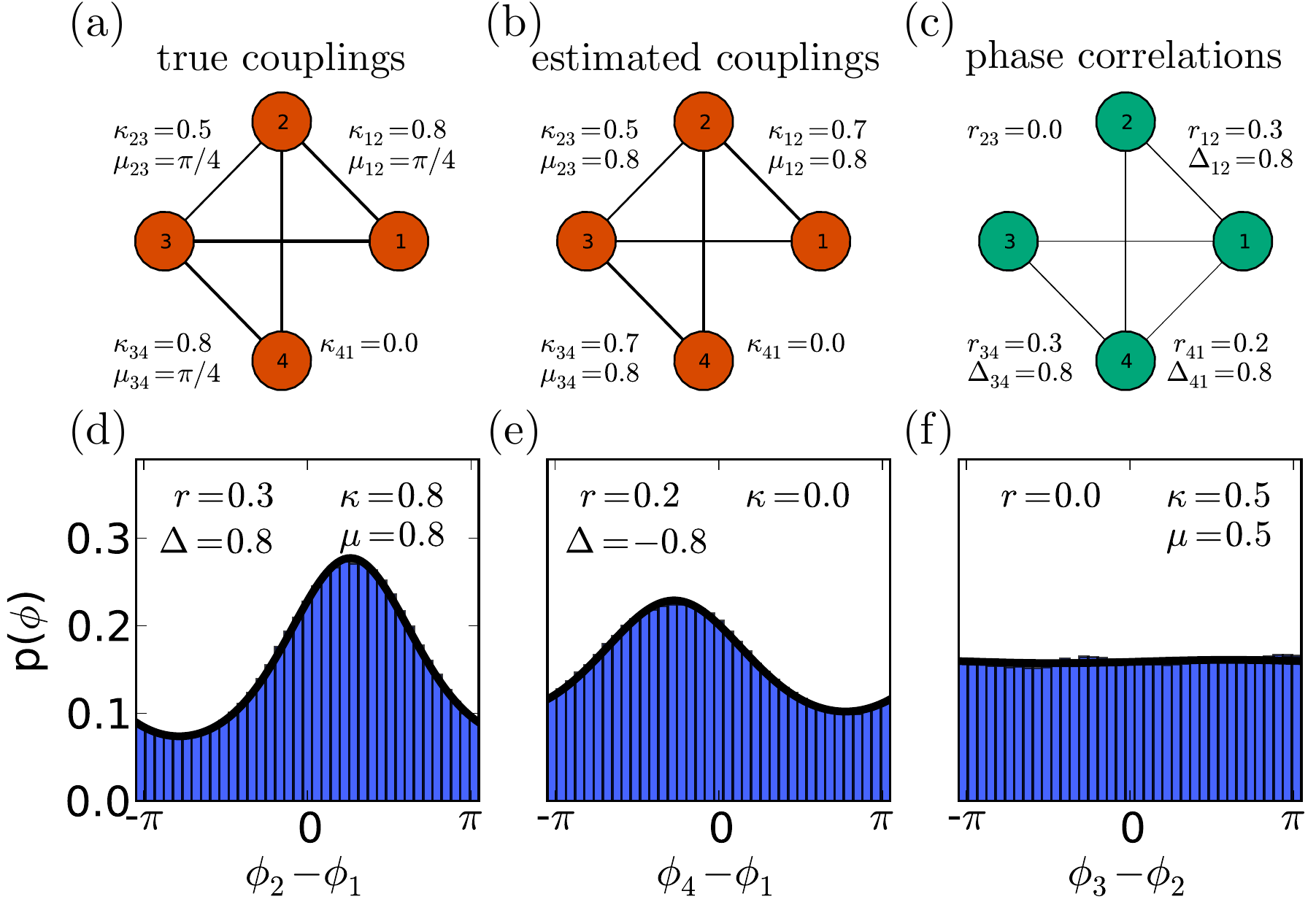}
\end{center}
\caption{Phase coupling estimation recovers the true coupling from
  measurements. (a)~Diagram of four coupled phase oscillators. (b)~Estimated
  coupling parameters using phase coupling estimation. (c)~Pair-wise phase
  correlations of the oscillators in a. Phase correlations are parameterized
  by the mean direction vector $r_{kl} e^{i\Delta_{kl}}\!=\!\langle
  e^{i\theta_l}e^{-i\theta_k}\rangle$ with amplitude $r_{kl}$ and angle
  $\Delta_{kl}$. (d)~Empirical phase distribution of the phase difference
  between oscillators 1 and 2. The empirical distribution is highly
  concentrated in the difference of the phases, while the marginals are flat
  (not shown). Phase correlations and estimated couplings are
  given. (e)~Distribution of phase differences of oscillators 1 and 4, which
  are not directly coupled. (f)~Uniform distribution of phase differences of
  oscillators 2 and 3, which are directly coupled. }
\label{fig:toy-model}
\end{figure}

In the following, we show that phase coupling estimation recovers the
parameters of simulated Kuramoto models. We begin by simulating a system of
four oscillators using equation~(\ref{eq:kuramoto}) with couplings shown in
Fig.~\ref{fig:toy-model}a. Given samples of the simulated phase variables
$\boldsymbol{\theta}$, we compute the correlations, $C_{ij}$, and required
four-point functions, $C_{ijkl}$, and invert the linear system in
equation~(\ref{eq:linearsystem}). This produces an estimate of the coupling
parameters. Phase coupling estimation recovers the true coupling parameters as
shown in Fig.~\ref{fig:toy-model}b.

We would like to point out that the pairwise statistics in systems of coupled
phase oscillators are only indirectly related to the coupling
parameters. Phase correlations, a pair-wise measurement often used to
characterize oscillator systems, have a direct relationship to the marginal
distribution of phase differences but not to coupling parameters. The form of
the marginal distribution can be derived by examination of the individual
factors in the definition
\begin{equation}
p(\theta_k\!-\!\theta_l) \sim \int \prod_{i,j=1}^d  \exp\left[
  \kappa_{ij}\cos(\theta_i-\theta_j-\mu_{ij}) \right] \mathrm{d}\theta^{d-2}\,,
\label{eq:marginals2d}
\end{equation}
in which the integration is over all phases $\theta_i$ with
$i\!\neq\!k,l$. After applying the variable substitution
$\tilde{\theta}_i\!=\!\theta_i+\theta_l$, all terms in~(\ref{eq:marginals2d})
either depend on the phase difference $\theta_k\!-\!\theta_l$, or are
independent of $\theta_k$ and $\theta_l$.  The independent terms integrate to
a constant and the remaining terms combine to a von Mises distribution in the
pairwise phase difference
\begin{equation}
  p(\theta_k\!-\!\theta_l) = \frac{1}{Z(\gamma_{kl})}
  e^{\displaystyle\gamma_{kl}\cos(\theta_k\!-\!\theta_l\!-\!\Delta_{kl})},
\label{eq:distribution2d}
\end{equation}
with mean phase $\Delta_{kl}$ and concentration parameter $\gamma_{kl}$. The
concentration parameter $\gamma_{kl}$ can be obtained by numerically solving
the equation $r_{kl}\!=\!I_1(\gamma_{kl})/I_0(\gamma_{kl})$ and the
normalization constant $Z(\gamma_{kl})$ is given by $Z(\gamma_{kl})\!=\!2\,\pi
I_0(\gamma_{kl})$. $I_0(x)$ and $I_1(x)$ denote the modified Bessel functions
of zeroth and first order, respectively. The value of $\gamma_{kl}$ is related
to the coupling parameters $\mathbf{K}$ through
equation~(\ref{eq:marginals2d}). Therefore, there is a non-trivial
relationship between the measured phase correlations and the coupling
parameters.

Because of this non-trivial relationship, pair-wise measurements can often
lead to false interpretations of the true coupling. We show the measured phase
correlations of our four oscillator system in
Fig.~\ref{fig:toy-model}c. Phases $\theta_1$ and $\theta_4$ show clear
correlations even though they are not directly coupled to each other (see
Fig.~\ref{fig:toy-model}e). Conversely, phases $\theta_2$ and $\theta_3$ are
strongly coupled but are uncorrelated (see Fig.~\ref{fig:toy-model}f).

We now systematically analyze the performance of phase coupling estimation:
the ability of the technique to recover the distribution parameters from
data. The procedure is as follows. We begin by sampling a set of distribution
parameters $\mathbf{K}$. Given these parameters we then sample phase variables
by numerically integrating~(\ref{eq:kuramoto}). We then estimate the
parameters given the sampled data using equation~(\ref{eq:linearsystem}). The
real and imaginary entries of the complex matrix $\mathbf{K}$ are sampled from
a normal distribution: $\mathrm{Re} \{K_{ij} \}, \mathrm{Im} \{K_{ij} \} \sim
N(0,1)$ and the diagonal entries are set to zero: $K_{ii} = 0$. Note that this
produces a dense coupling matrix.

\begin{figure}[ht]
\begin{center}
\includegraphics[width=8.5cm]{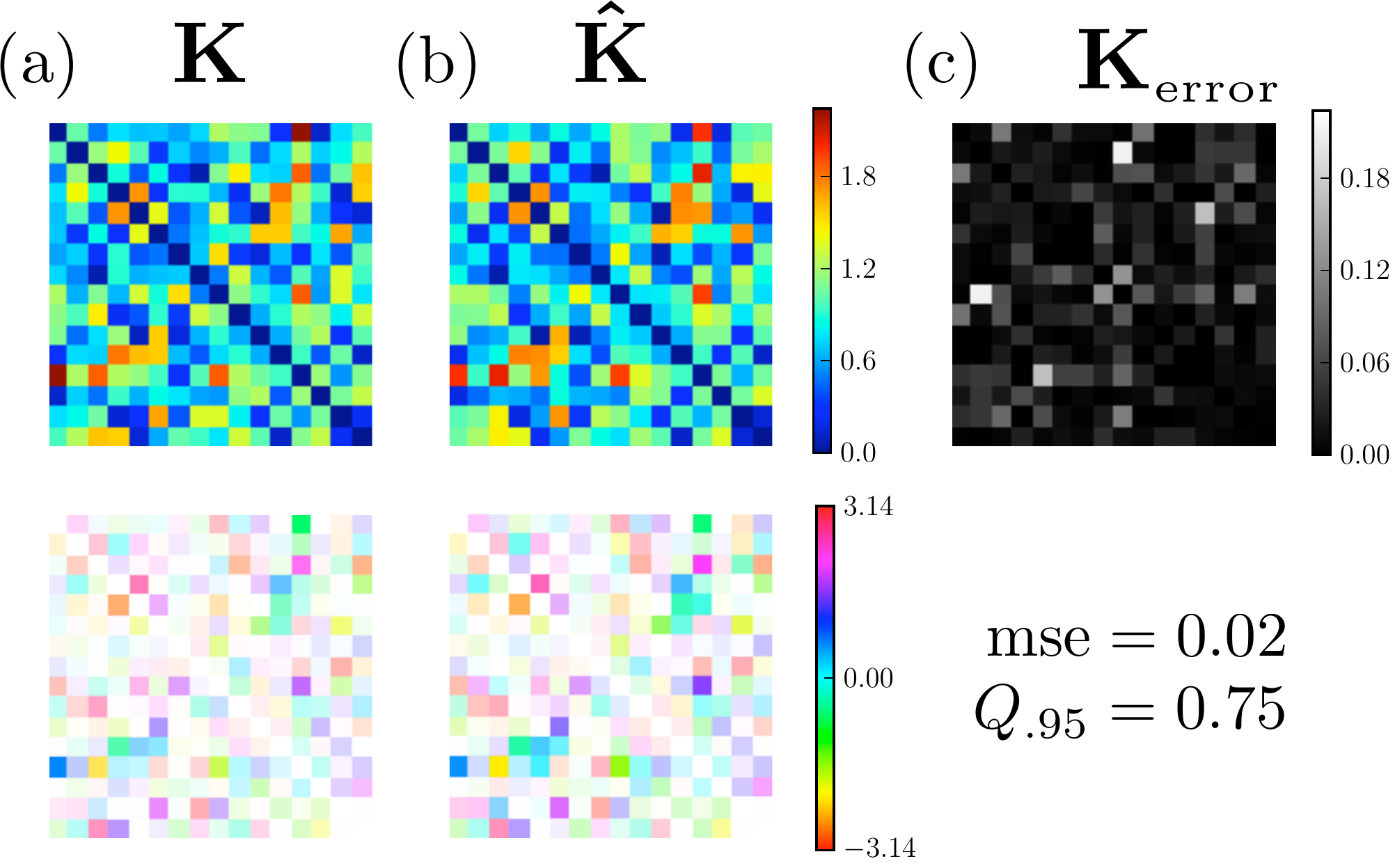} 
\end{center}
\caption{Phase coupling estimation for a system of 16 coupled oscillators with
  random coupling. (a)~True coupling matrix $\mathbf{K}$: true system
  parameters for $d=16$ (first row, element-wise amplitude; second row,
  element-wise angle with alpha channel scaled by the amplitude of the
  corresponding element, best viewed in color). (b)~estimated coupling matrix
  $\mathbf{\hat K }$: estimated parameters recovered from $2560$ time samples
  of $\boldsymbol{\theta}$ using
  equation~(\ref{eq:linearsystem}). (c)~estimation error: first row element-wise
  $\text{mse}$ (note scaling), second row estimation error measurements,
  $\text{mse}$ and $Q_{.95}$ (see text for definition). }
\label{fig:performance1}
\end{figure}

In the first column of Fig.~\ref{fig:performance1}, we graphically display the
element-wise amplitude and phase of a sample matrix $\mathbf{K}$ where
$d\!=\!16$. Using this matrix we sampled $N\!=\!2560$ phase vectors. The
recovered parameters are shown in the second column of
Fig.~\ref{fig:performance1}. While it is clear that these matrices are
visually similar, we quantified the error using two different metrics. First
we calculate the mean-squared-error of the matrix elements,
$\text{mse}\!=\!\frac{1}{2d^2}\sum_{i,j}{ | K_{ij}\!-\!\hat K_{ij} |^2 }$,
where $\hat K_{ij}$ is the estimated parameter. In the third column of
Fig.~\ref{fig:performance1} we display the element-wise error before
averaging. We also computed a metric indicating the quality of the recovered
parameters borrowed from Ref.~\cite{Timme2007}:
$Q_{.95}\!=\!\frac{1}{d^2}\sum_{i,j} u(1\!-.95 \Delta K_{ij})$, where $\Delta
K_{ij} = | K_{ij} - \hat K_{ij}|/2K_{\max} $, $u$ is the Heaviside step
function, and $K_{\max}$ is the maximum absolute value of all matrix entries
$K_{ij}$ and $\hat{K}_{ij}$. For the example in Fig.~\ref{fig:performance1},
$\text{mse}\!=\!0.02$, and $Q_{.95} = 0.75$.

\begin{figure}[ht]
\begin{center}
\includegraphics[width=8.5cm]{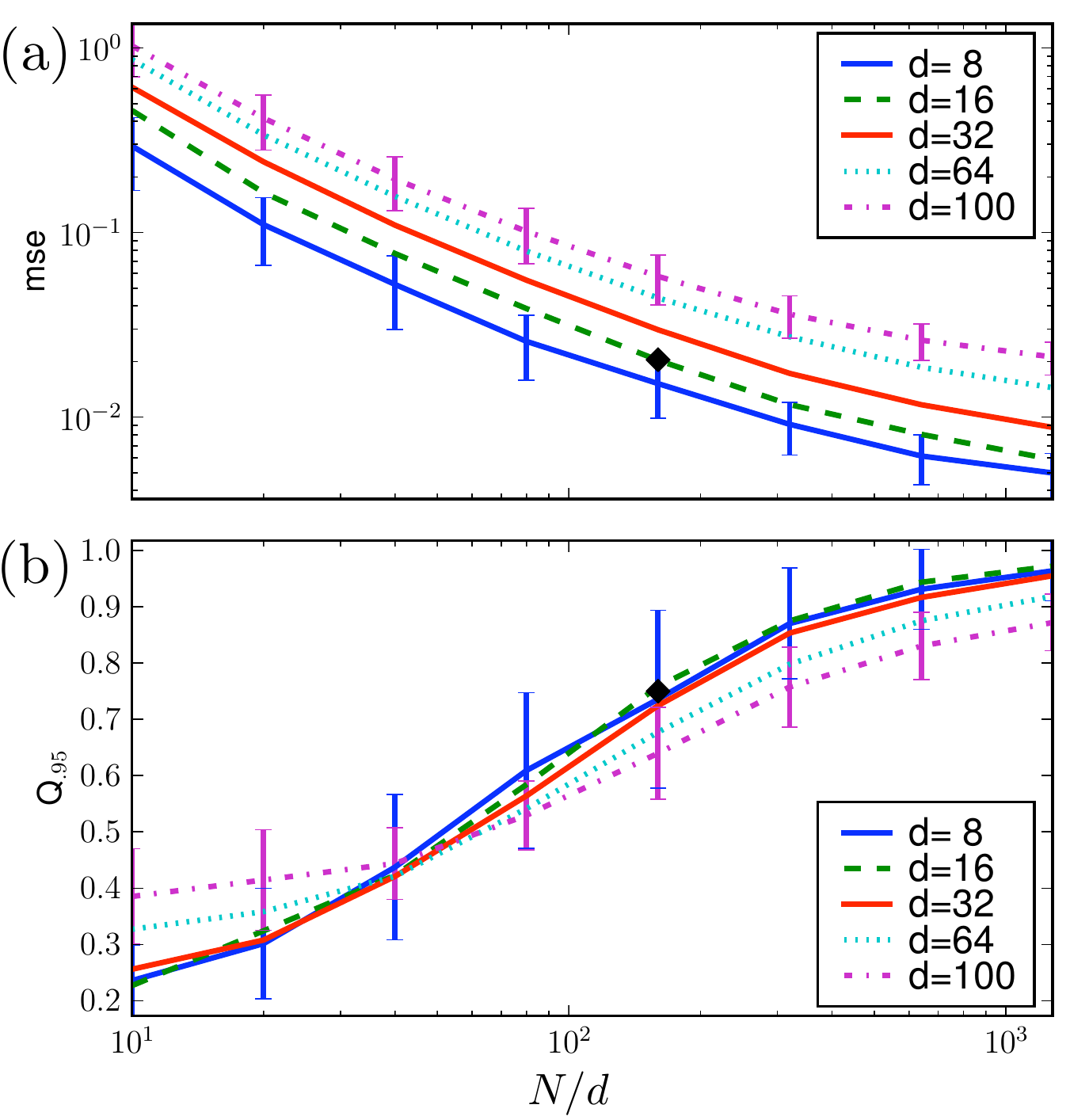} 
\end{center}
\caption{Performance of phase coupling estimation. (a) mean-squared-error,
  ~$\text{mse}$, metric as a function of samples per dimension for various
  dimensions, $d = 8,16,32,64,100$. (b)~$Q_{.95}$ metric. The example
  displayed in Fig.~\ref{fig:performance1} is indicated by the black
  diamond. Values are averaged over 20 trials. For visual clarity, standard
  error bars are only displayed for $d = 8$ and $d = 100$. Phase coupling
  estimation accurately recovers the system parameters with only 100 samples
  per dimension and achieves nearly perfect recovery with 1000 samples per
  dimension.}
\label{fig:performance2}
\end{figure}

We computed these error metrics over a range of dimensions and samples per
dimension. The error metrics for each dimension and samples per dimension were
averaged over 20 trials and are plotted in Fig.~\ref{fig:performance2}. The
algorithm achieves highly accurate parameter recovery for as few as 100
samples per dimension and achieves full recovery of parameters as the number
of samples per dimension reaches 1000. This indicates that recovery of true
parameters is quite feasible in many real world settings.

In this Letter we have introduced a closed form solution to the inverse
problem for systems of coupled phase oscillators using a maximum entropy
approach. We close by pointing out the relation of our work to other attempts
at solving the inverse problem for coupled phase oscillators. Previous
examples of probabilistic distributions of phase variables only characterize
univariate or bivariate distributions (see
e.g.~\cite{Jammalamadaka2001,Mardia2000}). Distributions similar to
equation~(\ref{eq:distribution}) have not been extended to dimensions beyond
$d\!=\!2$~\cite{Gatto2007,Mardia2007}. Common multivariate phase distributions
do not capture the statistics that are relevant for coupled phase
oscillators. Most notably the von Mises-Fisher distribution only captures a
unimodal phase distribution on the hyper-sphere, which does not produce
unimodal concentrations in the differences of phase variables. One of the most
relevant proposals is the estimation procedure of Timme~\cite{Timme2007}.
While the method of Ref.~\cite{Timme2007} successfully recovers the coupling
parameters, it requires repeated intervention, which may not be feasible in
many real-world experiments.

Phase coupling estimation can potentially provide a contribution in a variety
of situations of scientific interest. Because our phase coupled estimation
technique derives the unique maximum entropy solution, it serves as the least
biased estimate of the system possible and can be used when the true dynamics
of the system are unknown. Such situations are prevalent in neuroscience where
oscillations are thought to mitigate cognitive processes but the form of the
underlying dynamical system is unknown. This field has lacked a suitable
procedure for estimating phase interactions and has largely relied on
pair-wise statistical measurements (see e.g.~\cite{Varela2001}).

\begin{acknowledgments}
  We would like to thank Matthias Bethge, Chris Hillar, Martin Lisewski, Bruno
  Olshausen, Jascha Sohl-Dickstein and Mark Timme for helpful comments and
  discussions. This work has been supported by NSF grants IIS-0705939 and
  IIS-0713657, NGA grant HM1582-05-1-2017.
\end{acknowledgments}

\vspace*{-.5cm}

\begin{thebibliography}{19}
\vspace*{-.5cm}
\expandafter\ifx\csname natexlab\endcsname\relax\def\natexlab#1{#1}\fi
\expandafter\ifx\csname bibnamefont\endcsname\relax
  \def\bibnamefont#1{#1}\fi
\expandafter\ifx\csname bibfnamefont\endcsname\relax
  \def\bibfnamefont#1{#1}\fi
\expandafter\ifx\csname citenamefont\endcsname\relax
  \def\citenamefont#1{#1}\fi
\expandafter\ifx\csname url\endcsname\relax
  \def\url#1{\texttt{#1}}\fi
\expandafter\ifx\csname urlprefix\endcsname\relax\def\urlprefix{URL }\fi
\providecommand{\bibinfo}[2]{#2}
\providecommand{\eprint}[2][]{\url{#2}}

\bibitem[{\citenamefont{Winfree}(2001)}]{Winfree2001}
\bibinfo{author}{\bibfnamefont{A.}~\bibnamefont{Winfree}},
  \emph{\bibinfo{title}{{The Geometry of Biological Time}}}
  (\bibinfo{publisher}{Springer-Verlag, Berlin}, \bibinfo{year}{2001}).

\bibitem[{\citenamefont{Kuramoto}(1984)}]{Kuramoto1984}
\bibinfo{author}{\bibfnamefont{Y.}~\bibnamefont{Kuramoto}},
  \emph{\bibinfo{title}{Chemical Oscillations, Waves, and Turbulence}}
  (\bibinfo{publisher}{Springer-Verlag, Berlin}, \bibinfo{year}{1984}).

\bibitem[{\citenamefont{Mirollo and Strogatz}(1990)}]{Mirollo1990}
\bibinfo{author}{\bibfnamefont{R.}~\bibnamefont{Mirollo}} \bibnamefont{and}
  \bibinfo{author}{\bibfnamefont{S.}~\bibnamefont{Strogatz}},
  \bibinfo{journal}{SIAM J. Appl. Math} \textbf{\bibinfo{volume}{50}},
  \bibinfo{pages}{1645} (\bibinfo{year}{1990}).

\bibitem[{\citenamefont{Strogatz}(2003)}]{Strogatz2003}
\bibinfo{author}{\bibfnamefont{S.}~\bibnamefont{Strogatz}},
  \emph{\bibinfo{title}{{Sync: The Emerging Science of Spontaneous Order}}}
  (\bibinfo{publisher}{Hyperion}, \bibinfo{year}{2003}).

\bibitem[{\citenamefont{Buzsaki}(2006)}]{Buzsaki2006}
\bibinfo{author}{\bibfnamefont{G.}~\bibnamefont{Buzsaki}},
  \emph{\bibinfo{title}{{Rhythms of the Brain}}} (\bibinfo{publisher}{Oxford
  University Press, USA}, \bibinfo{year}{2006}).

\bibitem[{\citenamefont{Jaynes}(1957)}]{Jaynes1957}
\bibinfo{author}{\bibfnamefont{E.}~\bibnamefont{Jaynes}},
  \bibinfo{journal}{Physical review} \textbf{\bibinfo{volume}{106}},
  \bibinfo{pages}{620} (\bibinfo{year}{1957}).

\bibitem[{\citenamefont{Shlens et~al.}(2006)\citenamefont{Shlens, Field,
  Gauthier, Grivich, Petrusca, Sher, Litke, and Chichilnisky}}]{Shlens2006}
\bibinfo{author}{\bibfnamefont{J.}~\bibnamefont{Shlens}},
  \bibinfo{author}{\bibfnamefont{G.~D.} \bibnamefont{Field}},
  \bibinfo{author}{\bibfnamefont{J.~L.} \bibnamefont{Gauthier}},
  \bibinfo{author}{\bibfnamefont{M.~I.} \bibnamefont{Grivich}},
  \bibinfo{author}{\bibfnamefont{D.}~\bibnamefont{Petrusca}},
  \bibinfo{author}{\bibfnamefont{A.}~\bibnamefont{Sher}},
  \bibinfo{author}{\bibfnamefont{A.~M.} \bibnamefont{Litke}}, \bibnamefont{and}
  \bibinfo{author}{\bibfnamefont{E.~J.} \bibnamefont{Chichilnisky}},
  \bibinfo{journal}{J Neurosci} \textbf{\bibinfo{volume}{26}},
  \bibinfo{pages}{8254} (\bibinfo{year}{2006}).

\bibitem[{\citenamefont{Schneidman et~al.}(2006)\citenamefont{Schneidman,
  Berry, Segev, and Bialek}}]{Schneidman2006}
\bibinfo{author}{\bibfnamefont{E.}~\bibnamefont{Schneidman}},
  \bibinfo{author}{\bibfnamefont{M.~J.} \bibnamefont{Berry}},
  \bibinfo{author}{\bibfnamefont{R.}~\bibnamefont{Segev}}, \bibnamefont{and}
  \bibinfo{author}{\bibfnamefont{W.}~\bibnamefont{Bialek}},
  \bibinfo{journal}{Nature} \textbf{\bibinfo{volume}{440}},
  \bibinfo{pages}{1007} (\bibinfo{year}{2006}).

\bibitem[{\citenamefont{Risken}(1989)}]{Risken1989a}
\bibinfo{author}{\bibfnamefont{H.}~\bibnamefont{Risken}},
  \emph{\bibinfo{title}{{The Fokker-Planck equation: Methods of solution and
  applications}}} (\bibinfo{publisher}{Springer-Verlag, Berlin},
  \bibinfo{year}{1989}).

\bibitem[{\citenamefont{Hyv{\"a}rinen}(2005)}]{Hyvarinen2005}
\bibinfo{author}{\bibfnamefont{A.}~\bibnamefont{Hyv{\"a}rinen}},
  \bibinfo{journal}{The Journal of Machine Learning Research}
  \textbf{\bibinfo{volume}{6}}, \bibinfo{pages}{695} (\bibinfo{year}{2005}).

\bibitem[{\citenamefont{Hyv{\"a}rinen}(2007)}]{Hyvarinen2007}
\bibinfo{author}{\bibfnamefont{A.}~\bibnamefont{Hyv{\"a}rinen}},
  \bibinfo{journal}{Computational Statistics and Data Analysis}
  \textbf{\bibinfo{volume}{51}}, \bibinfo{pages}{2499} (\bibinfo{year}{2007}).



\bibitem[{\citenamefont{Timme}(2007)}]{Timme2007}
\bibinfo{author}{\bibfnamefont{M.}~\bibnamefont{Timme}},
  \bibinfo{journal}{Physical Review Letters} \textbf{\bibinfo{volume}{98}},
  \bibinfo{pages}{224101} (\bibinfo{year}{2007}).

\bibitem[{\citenamefont{Jammalamadaka and Sengupta}(2001)}]{Jammalamadaka2001}
\bibinfo{author}{\bibfnamefont{S.}~\bibnamefont{Jammalamadaka}}
  \bibnamefont{and} \bibinfo{author}{\bibfnamefont{A.}~\bibnamefont{Sengupta}},
  \emph{\bibinfo{title}{{Topics in Circular Statistics}}}
  (\bibinfo{publisher}{World Scientific}, \bibinfo{year}{2001}).

\bibitem[{\citenamefont{Mardia and Jupp}(2000)}]{Mardia2000}
\bibinfo{author}{\bibfnamefont{K.}~\bibnamefont{Mardia}} \bibnamefont{and}
  \bibinfo{author}{\bibfnamefont{P.}~\bibnamefont{Jupp}},
  \emph{\bibinfo{title}{{Directional statistics}}} (\bibinfo{publisher}{Wiley,
  New York}, \bibinfo{year}{2000}).

\bibitem[{\citenamefont{Gatto and Jammalamadaka}(2007)}]{Gatto2007}
\bibinfo{author}{\bibfnamefont{R.}~\bibnamefont{Gatto}} \bibnamefont{and}
  \bibinfo{author}{\bibfnamefont{S.}~\bibnamefont{Jammalamadaka}},
  \bibinfo{journal}{Statistical Methodology} \textbf{\bibinfo{volume}{4}},
  \bibinfo{pages}{341} (\bibinfo{year}{2007}).

\bibitem[{\citenamefont{Mardia et~al.}(2007)\citenamefont{Mardia, Taylor, and
  Subramaniam}}]{Mardia2007}
\bibinfo{author}{\bibfnamefont{K.}~\bibnamefont{Mardia}},
  \bibinfo{author}{\bibfnamefont{C.}~\bibnamefont{Taylor}}, \bibnamefont{and}
  \bibinfo{author}{\bibfnamefont{G.}~\bibnamefont{Subramaniam}},
  \bibinfo{journal}{Biometrics} \textbf{\bibinfo{volume}{63}},
  \bibinfo{pages}{505} (\bibinfo{year}{2007}).

\bibitem[{\citenamefont{Varela et~al.}(2001)\citenamefont{Varela, Lachaux,
  Rodriguez, and Martinerie}}]{Varela2001}
\bibinfo{author}{\bibfnamefont{F.}~\bibnamefont{Varela}},
  \bibinfo{author}{\bibfnamefont{J.}~\bibnamefont{Lachaux}},
  \bibinfo{author}{\bibfnamefont{E.}~\bibnamefont{Rodriguez}},
  \bibnamefont{and}
  \bibinfo{author}{\bibfnamefont{J.}~\bibnamefont{Martinerie}},
  \bibinfo{journal}{Nature Reviews Neuroscience} \textbf{\bibinfo{volume}{2}},
  \bibinfo{pages}{229} (\bibinfo{year}{2001}).

\end{thebibliography}

\end{document}